\documentclass{elsart_sf}
\usepackage{epsfig}
\usepackage{amssymb, amsmath}
\RequirePackage{lineno}
\begin{document}
\begin{frontmatter}

%\linenumbers


\begin{center}
\title{The ANTARES Telescope Neutrino Alert System}
\end{center}
\author[CPPM]{M.~Ageron},
\author[IFIC]{J.A.~Aguilar},
\author[CPPM]{I.~Al~Samarai},
\author[Colmar]{A.~Albert},
\author[Barcelona]{M.~Andr\'e},
\author[Genova]{M.~Anghinolfi},
\author[Erlangen]{G.~Anton},
\author[IRFU/SEDI]{S.~Anvar},
\author[UPV]{M.~Ardid},
\author[NIKHEF]{A.C.~Assis Jesus},
% ...

%\newpage
\nopagebreak[3]
\address[CPPM]{\scriptsize{CPPM, Aix-Marseille Universit\'e, CNRS/IN2P3, Marseille, France}}\vspace*{0.15cm}
\nopagebreak[3]
\vspace*{-0.20\baselineskip}
\nopagebreak[3]
\address[IFIC]{\scriptsize{IFIC - Instituto de F\'isica Corpuscular, Edificios Investigaci\'on de Paterna, CSIC - Universitat de Val\`encia, Apdo. de Correos 22085, 46071 Valencia, Spain}}\vspace*{0.15cm}
\nopagebreak[3]
\vspace*{-0.20\baselineskip}
\nopagebreak[3]
\address[Colmar]{\scriptsize{GRPHE - Institut universitaire de technologie de Colmar, 34 rue du Grillenbreit BP 50568 - 68008 Colmar, France }}\vspace*{0.15cm}
\nopagebreak[3]
\vspace*{-0.20\baselineskip}
\nopagebreak[3]
\address[Barcelona]{\scriptsize{Technical University of Catalonia, Laboratory of Applied Bioacoustics, Rambla Exposici\'o,08800 Vilanova i la Geltr\'u,Barcelona, Spain}}\vspace*{0.15cm}
\nopagebreak[3]
\vspace*{-0.20\baselineskip}
\nopagebreak[3]
\address[Genova]{\scriptsize{INFN - Sezione di Genova, Via Dodecaneso 33, 16146 Genova, Italy}}\vspace*{0.15cm}
\nopagebreak[3]
\vspace*{-0.20\baselineskip}
\nopagebreak[3]
\address[Erlangen]{\scriptsize{Friedrich-Alexander-Universit\"at Erlangen-N\"urnberg, Erlangen Centre for Astroparticle Physics, Erwin-Rommel-Str. 1, 91058 Erlangen, Germany}}\vspace*{0.15cm}
\nopagebreak[3]
\vspace*{-0.20\baselineskip}
\nopagebreak[3]
\address[IRFU/SEDI]{\scriptsize{Direction des Sciences de la Mati\`ere - Institut de recherche sur les lois fondamentales de l'Univers - Service d'Electronique des D\'etecteurs et d'Informatique, CEA Saclay, 91191 Gif-sur-Yvette Cedex, France}}\vspace*{0.15cm}
\nopagebreak[3]
\vspace*{-0.20\baselineskip}
\nopagebreak[3]
\address[UPV]{\scriptsize{Institut d'Investigaci\'o per a la Gesti\'o Integrada de les Zones Costaneres (IGIC) - Universitat Polit\`ecnica de Val\`encia. C/  Paranimf 1 , 46730 Gandia, Spain.}}\vspace*{0.15cm}
\nopagebreak[3]
\vspace*{-0.20\baselineskip}
\nopagebreak[3]
\address[NIKHEF]{\scriptsize{Nikhef, Science Park,  Amsterdam, The Netherlands}}\vspace*{0.15cm}
\nopagebreak[3]
\vspace*{-0.20\baselineskip}
\nopagebreak[3]
\address[APC]{\scriptsize{APC - Laboratoire AstroParticule et Cosmologie, UMR 7164 (CNRS, Universit\'e Paris 7 Diderot, CEA, Observatoire de Paris) 10, rue Alice Domon et L\'eonie Duquet 75205 Paris Cedex 13,  France}}\vspace*{0.15cm}
\nopagebreak[3]
\vspace*{-0.20\baselineskip}
\nopagebreak[3]
\address[LAM]{\scriptsize{LAM - Laboratoire d'Astrophysique de Marseille, P\^ole de l'\'Etoile Site de Ch\^ateau-Gombert, rue Fr\'ed\'eric Joliot-Curie 38,  13388 Marseille Cedex 13, France }}\vspace*{0.15cm}
\nopagebreak[3]
\vspace*{-0.20\baselineskip}
\nopagebreak[3]
\address[Bologna]{\scriptsize{INFN - Sezione di Bologna, Viale Berti Pichat 6/2, 40127 Bologna, Italy}}\vspace*{0.15cm}
\nopagebreak[3]
\vspace*{-0.20\baselineskip}
\nopagebreak[3]
\address[Bologna-UNI]{\scriptsize{Dipartimento di Fisica dell'Universit\`a, Viale Berti Pichat 6/2, 40127 Bologna, Italy}}\vspace*{0.15cm}
\nopagebreak[3]
\vspace*{-0.20\baselineskip}
\nopagebreak[3]
\address[Pisa]{\scriptsize{INFN - Sezione di Pisa, Largo B. Pontecorvo 3, 56127 Pisa, Italy}}\vspace*{0.15cm}
\nopagebreak[3]
\vspace*{-0.20\baselineskip}
\nopagebreak[3]
\address[Roma]{\scriptsize{INFN -Sezione di Roma, P.le Aldo Moro 2, 00185 Roma, Italy}}\vspace*{0.15cm}
\nopagebreak[3]
\vspace*{-0.20\baselineskip}
\nopagebreak[3]
\address[Roma-UNI]{\scriptsize{Dipartimento di Fisica dell'Universit\`a La Sapienza, P.le Aldo Moro 2, 00185 Roma, Italy}}\vspace*{0.15cm}
\nopagebreak[3]
\vspace*{-0.20\baselineskip}
\nopagebreak[3]
\address[Clermont-Ferrand]{\scriptsize{Clermont Universit\'e, Universit\'e Blaise Pascal, CNRS/IN2P3, Laboratoire de Physique Corpusculaire, BP 10448, 63000 Clermont-Ferrand, France}}\vspace*{0.15cm}
\nopagebreak[3]
\vspace*{-0.20\baselineskip}
\nopagebreak[3]
\address[INAF]{\scriptsize{INAF-IASF, via P. Gobetti 101, 40129 Bologna, Italy}}\vspace*{0.15cm}
\nopagebreak[3]
\vspace*{-0.20\baselineskip}
\nopagebreak[3]
\address[GEOAZUR]{\scriptsize{G\'eoazur - Universit\'e de Nice Sophia-Antipolis, CNRS/INSU, IRD, Observatoire de la C\^ote d'Azur and Universit\'e Pierre et Marie Curie, BP 48, 06235 Villefranche-sur-mer, France}}\vspace*{0.15cm}
\nopagebreak[3]
\vspace*{-0.20\baselineskip}
\nopagebreak[3]
\address[Bari]{\scriptsize{INFN - Sezione di Bari, Via E. Orabona 4, 70126 Bari, Italy}}\vspace*{0.15cm}
\nopagebreak[3]
\vspace*{-0.20\baselineskip}
\nopagebreak[3]
\address[COM]{\scriptsize{COM - Centre d'Oc\'eanologie de Marseille, CNRS/INSU et Universit\'e de la M\'editerran\'ee, 163 Avenue de Luminy, Case 901, 13288 Marseille Cedex 9, France}}\vspace*{0.15cm}
\nopagebreak[3]
\vspace*{-0.20\baselineskip}
\nopagebreak[3]
\address[LNS]{\scriptsize{INFN - Laboratori Nazionali del Sud (LNS), Via S. Sofia 62, 95123 Catania, Italy}}\vspace*{0.15cm}
\nopagebreak[3]
\vspace*{-0.20\baselineskip}
\nopagebreak[3]
\address[UPS]{\scriptsize{Univ Paris-Sud , 91405 Orsay Cedex, France}}\vspace*{0.15cm}
\nopagebreak[3]
\vspace*{-0.20\baselineskip}
\nopagebreak[3]
\address[KVI]{\scriptsize{Kernfysisch Versneller Instituut (KVI), University of Groningen, Zernikelaan 25, 9747 AA Groningen, The Netherlands}}\vspace*{0.15cm}
\nopagebreak[3]
\vspace*{-0.20\baselineskip}
\nopagebreak[3]
\address[Pisa-UNI]{\scriptsize{Dipartimento di Fisica dell'Universit\`a, Largo B. Pontecorvo 3, 56127 Pisa, Italy}}\vspace*{0.15cm}
\nopagebreak[3]
\vspace*{-0.20\baselineskip}
\nopagebreak[3]
\address[NIOZ]{\scriptsize{Royal Netherlands Institute for Sea Research (NIOZ), Landsdiep 4,1797 SZ 't Horntje (Texel), The Netherlands}}\vspace*{0.15cm}
\nopagebreak[3]
\vspace*{-0.20\baselineskip}
\nopagebreak[3]
\address[Bamberg]{\scriptsize{Dr. Remeis-Sternwarte and ECAP, Universit\"at Erlangen-N\"urnberg,  Sternwartstr. 7, 96049 Bamberg, Germany}}\vspace*{0.15cm}
\nopagebreak[3]
\vspace*{-0.20\baselineskip}
\nopagebreak[3]
\address[UU]{\scriptsize{Universiteit Utrecht, Faculteit Betawetenschappen, Princetonplein 5, 3584 CC Utrecht, The Netherlands}}\vspace*{0.15cm}
\nopagebreak[3]
\vspace*{-0.20\baselineskip}
\nopagebreak[3]
\address[UvA]{\scriptsize{Universiteit van Amsterdam, Instituut voor Hoge-Energie Fysika, Science Park 105, 1098 XG Amsterdam, The Netherlands}}\vspace*{0.15cm}
\nopagebreak[3]
\vspace*{-0.20\baselineskip}
\nopagebreak[3]
\address[MSU]{\scriptsize{Moscow State University,Skobeltsyn Institute of Nuclear Physics,Leninskie gory, 119991 Moscow, Russia}}\vspace*{0.15cm}
\nopagebreak[3]
\vspace*{-0.20\baselineskip}
\nopagebreak[3]
\address[Catania]{\scriptsize{INFN - Sezione di Catania, Viale Andrea Doria 6, 95125 Catania, Italy}}\vspace*{0.15cm}
\nopagebreak[3]
\vspace*{-0.20\baselineskip}
\nopagebreak[3]
\address[Catania-UNI]{\scriptsize{Dipartimento di Fisica ed Astronomia dell'Universit\`a, Viale Andrea Doria 6, 95125 Catania, Italy}}\vspace*{0.15cm}
\nopagebreak[3]
\vspace*{-0.20\baselineskip}
\nopagebreak[3]
\address[IRFU/SPP]{\scriptsize{Direction des Sciences de la Mati\`ere - Institut de recherche sur les lois fondamentales de l'Univers - Service de Physique des Particules, CEA Saclay, 91191 Gif-sur-Yvette Cedex, France}}\vspace*{0.15cm}
\nopagebreak[3]
\vspace*{-0.20\baselineskip}
\nopagebreak[3]
\address[WIN]{\scriptsize{University of Wisconsin - Madison, 53715, WI, USA}}\vspace*{0.15cm}
\nopagebreak[3]
\vspace*{-0.20\baselineskip}
\nopagebreak[3]
\address[ISS]{\scriptsize{Institute for Space Sciences, R-77125 Bucharest, M\u{a}gurele, Romania     }}\vspace*{0.15cm}
\nopagebreak[3]
\vspace*{-0.20\baselineskip}
\nopagebreak[3]
\address[IPHC]{\scriptsize{IPHC-Institut Pluridisciplinaire Hubert Curien - Universit\'e de Strasbourg et CNRS/IN2P3  23 rue du Loess, BP 28,  67037 Strasbourg Cedex 2, France}}\vspace*{0.15cm}
\nopagebreak[3]
\vspace*{-0.20\baselineskip}
\nopagebreak[3]
\address[ITEP]{\scriptsize{ITEP - Institute for Theoretical and Experimental Physics, B. Cheremushkinskaya 25, 117218 Moscow, Russia}}\vspace*{0.15cm}
\nopagebreak[3]
\vspace*{-0.20\baselineskip}
\nopagebreak[3]
\address[Genova-UNI]{\scriptsize{Dipartimento di Fisica dell'Universit\`a, Via Dodecaneso 33, 16146 Genova, Italy}}\vspace*{0.15cm}
\nopagebreak[3]
\vspace*{-0.20\baselineskip}
%\input{author_list_feb11}
%\begin{center}
%{\bf The ANTARES Collaboration}\\
%\end{center}

\begin{abstract}
{The ANTARES telescope has the capability
to detect neutrinos
produced in astrophysical
transient sources.
Potential sources include gamma-ray bursts,
core collapse supernovae, and
flaring active galactic nuclei.}
{To enhance the sensitivity of ANTARES to such
sources, a new detection method based on coincident observations of neutrinos
and optical signals has been developed.
A fast online muon track reconstruction is used to trigger a network of
small automatic optical telescopes.} 
{Such alerts are generated for special events,
such as two or more neutrinos, coincident in time and direction, or single
neutrinos of very high energy.}  {}
\end{abstract}

\end{frontmatter}
% comment next line for arXiv
%\linenumbers

%\setcounter{footnote}{0}

%\include{introduction}
\section{Introduction}
The detection of high energy cosmic neutrino from a source would be
direct evidence
of the presence of hadronic acceleration within the source and provide
important
information on the origin of the high energy cosmic rays. Powerful
sources of transient nature,
such as gamma ray bursts or core collapse supernovae, offer one of the
most promising
perspectives for the detection of cosmic neutrinos as, due to their
short duration, they
are essentially background free.
For example,
several authors predict the emission of neutrinos in correlation with
multi-wavelength signals, e.g. the Fireball model of
GRBs~\cite{bib:FireballRef}.
As neutrino telescopes observe a full
hemisphere of the sky
(even the whole sky if downgoing events are considered) at all times,
they are particularly well
suited for the detection of transient phenomena.
%The detection of high energy cosmic neutrinos would be a direct proof
%of the existence of hadronic acceleration processes in astrophysical objects,
%thereby identifying the origin of very high energy cosmic rays.
%Powerful sources of transient nature such as gamma-ray bursts (GRB) or
%core collapse supernovae (SNe) offer one of the most promising perspectives
%for the detection of cosmic neutrinos as they are essentially background free
%due to their short duration.
%Several authors predict the emission of neutrinos in correlation with other
%multi-wavelength signals, for example the Fireball model of
%GRBs~\cite{bib:FireballRef}.
%Neutrino telescopes are well suited to
%survey a full hemisphere at all times and
%even 4$\pi$~sr if downgoing events as well as upgoing events
%are also considered.

In this paper, 
the implementation of a
strategy for the detection of transient sources is presented.
This method, earlier proposed in~\cite{bib:Marek},
is based on the optical follow-up of selected neutrino events
very shortly after their
detection (Section~\ref{sec:TatooRC}) by the ANTARES neutrino
telescope~\cite{bib:Antares}.
The alert system,
known as ``TAToO'' (Telescopes
and ANTARES Target of Opportunity)~\cite{bib:vlvnt09},
uses an online track reconstruction with a
pointing accuracy of about 0.5 degrees.
This reconstruction algorithm is described in Section~\ref{sec:AlertCriteria}.
Its characteristics allow the triggering of optical telescopes such as
TAROT~\cite{bib:Tarot} and ROTSE~\cite{bib:Rotse}.
In order to improve the precision of these alerts, an additional reconstruction
algorithm, described in Section~\ref{sec:RefinedPosition},
which takes into account the detailed detector geometry is used offline.

\section{The ANTARES detector and data acquisition}
The ANTARES neutrino telescope is located in the
Mediterranean Sea, 40~km from the coast of Toulon, France, at a depth of
2475~m. The detector is an array of photomultiplier
tubes (PMTs) arranged on 12 slender detection lines,
anchored to the sea bed and kept taut by a buoy at the top.
Each line comprises up to 25 storeys of triplets of optical modules (OMs),
each housing a single 10" PMT.
Since lines are subject to the sea current and can change shape and orientation,
a positioning system comprising hydrophones and compass-tiltmeters
is used to monitor the detector geometry.
Data taking started in 2006 with the operation of the
first line of the detector.
The construction of the 12 line detector was completed in May 2008.
The main goal of the experiment is to search for neutrinos of astrophysical
origin by detecting high energy muons ($\geq$100~GeV) induced by their
neutrino charged current interaction in the vicinity of the detector.
Due to the large background from downgoing cosmic ray induced muons, the
detector is optimised for the detection of upgoing neutrino induced muon tracks.

\subsection{Data acquisition}
The task of the ANTARES data acquisition system (DAQ)~\cite{bib:antaresdaq}
is to collect the data from all
the individual PMTs of the detector and pass them to the filtering
algorithms which search for a collection of signals compatible
with a muon track crossing the detector.

From the DAQ point of view, each storey is an independent acquisition unit
including a processor, buffering RAM and an Ethernet link to the shore station.
Individual
PMT pulses above a threshold of typically 0.3 photoelectron (referred to
as `hits') are digitized offshore
in the form of `hit time' and `charge', and sent to a computer farm
onshore for further processing. Due to bioluminescence activity
and $^{40}$K decays in the sea water, each PMT has an
average counting rate of the
order of 100 kHz, requiring large bandwidth for data
transmission to shore. The data are transmitted through gigabit Ethernet links
on single optical fibers to the computing farm where the filtering algorithms
are executed, reducing the event rate to a few tens of hertz.
The overall system is supervised by a state machine which handles the
various commands needed to configure, start and stop the acquisition on
both the offshore and onshore processors.

Data are time structured in the form of time slices of 104.85~ms,
allowing the data of the full detector for the duration of one time slice to be
sent to a single computing node. The synchronization of the 300 offshore
processors is performed by a 20~MHz clock distribution system
broadcasted to all storeys.  In particular, the start of any data taking
period is stamped using an external GPS signal giving the absolute timing
at the location of the detector, allowing an absolute time accuracy
better than 1~$\mu$s~\cite{bib:TimeCalib}.

\subsection{Data filtering}
The main goal of the filtering algorithm is to select hits
compatible with the
propagation of Cherenkov light emitted by a muon crossing the detector,
among the background from bioluminescence and $^{40}$K decays characterized
by uncorrelated hits on single OMs.
The filtering is based on local `clusters', defined either as
coincidence hits on OMs of the same storey within a narrow time
window or as a single hit with a large amplitude. 
There are two main filtering algorithms
running simultaneously, both searching for a combination of local
clusters
within a typical 2.2~$\mu$s time window.  The first algorithm
requires five local causally connected clusters anywhere in the detector,
while the second requires at least two local clusters in
adjacent or next-to-adjacent storeys. 
All hits within a few microseconds around these clusters
define an
``event'' and are kept for further online and offline reconstructions.

\subsection{The TAToO trigger}
Once a single data filtering computer has processed its time
slice, the resulting events are sent to a data distribution service to which the
``Data Writer" storing task, monitoring tasks and the online event
reconstruction~\cite{bib:BBfit} are connected. 
The ``alert" application described in
Section~\ref{sec:AlertCriteria} analyses the data stream of reconstructed
events, selects candidates fulfilling various criteria, and then
generates the TAToO alerts.

An important performance parameter for the alert
system is the time between the crossing of the detector by a high energy
muon and the time at which an alert is sent.
This time is the sum of the data
dispatching time from the offshore photomultipliers to the onshore
computing farm (1.5~s)
and of the data processing time of
an entire time slice by the filtering algorithm ($\leq$5~s).
For the time-being, an extra buffering time of 60 seconds
is added; this delay will be removed
after an upgrade of the acquisition to improve the performance of the alert
sending capability.
The time needed to reconstruct the event direction and verify the
alert criteria typically amounts to a few milliseconds.
Therefore, the total delay between an interesting particle crossing the
detector and a TAToO alert is currently 66~s.

\section{The TAToO Run Control}
\label{sec:TatooRC}
The TAToO Run Control (RC) is a stand-alone {\it Qt}
control application\footnote{Software
framework originally from Trolltech, now Nokia.} which channels
the triggers generated by the alert
application to the optical telescope network.  

The connections to this network are checked periodically and automatic
reconnection is performed resulting in a fully autonomous and stable system. 
A veto prevents an alert to be sent if the ANTARES event counting rate exceeds
a given threshold. In addition, if the alert criteria are fulfilled soon after a
previous alert has already been issued, the new alert is stored in a FIFO and
sent only after a certain period of time. This time lag, currently set at one
hour, is used to avoid alert pileup in the optical telescope network.
Manual alerts can also be generated and sent.  All alerts are
sent using the Gamma-ray bursts Coordinates Network
(GCN)~\cite{bib:GCN} normalized format, allowing easy implementation
of connections to additional telescopes.
Information about the event that triggered the
alert, i.e. a unique identifier, the time and the celestial coordinates,
the number of hits used in the reconstruction
and the track reconstruction quality are sent to
the optical telescopes network at the time of the alert.

\section{The alert criteria}
\label{sec:AlertCriteria}
The criteria for the TAToO trigger are based on the features of the
neutrino signal from the expected sources.
Several models predict the production of high energy neutrinos greater than
1 TeV from GRBs~\cite{bib:GRB} and from Core Collapse
Supernovae~\cite{bib:CCSN}. Under certain conditions, multiplet of
neutrinos can be expected~\cite{bib:CCSN1}.

Two online neutrino trigger criteria are currently implemented in the TAToO
alert system:
\begin{itemize} 
\item the detection of at least two neutrino induced muons coming from
similar directions within a predefined time window;
\item the detection of a single high energy neutrino induced muon.
\end{itemize}

A basic requirement for the coincident observation of a neutrino and an optical
counterpart is that the pointing accuracy of the neutrino telescope should be
at least comparable to the field of view of the optical telescope.  

\subsection{The online track reconstruction algorithm}
\label{sec:OnlineReco}
To select the events which might trigger an alert, a fast and robust algorithm
is used online to reconstruct tracks from the calibrated data.
This algorithm uses an idealized detector
geometry which does not rely on the dynamical positioning alignment.
As a result, the hits of the three OMs of a storey are grouped and
their location assigned to the barycenter of the storey.
The storey
orientations as well as the line-shape deviations from straight lines are
not considered in the online reconstruction.
A detailed description of this algorithm and its performance
is found in Ref.~\cite{bib:BBfit}.
The principle is to minimize a $\chi^2$ which compares the times of selected
hits with the expectation from a Cherenkov signal generated by a muon track.
The resulting direction of the reconstructed muon track is available within
$10~\mathrm{ms}$ and the obtained minimum $\chi^2$ is used as a fit quality
parameter to remove badly reconstructed tracks.

\subsection{Neutrino selection criteria}
Atmospheric muons, whose abundance at the ANTARES detector~\cite{bib:atmu}
is roughly six orders of magnitude larger than the one
of muons induced by atmospheric
neutrinos, are the main background for the alerts and have to be
efficiently suppressed. Among the surviving events, neutrino candidates
with an increased probability to be of cosmic origin are
selected~\cite{bib:difflux}.

\subsubsection{Atmospheric muon background rejection}
Atmospheric muons resulting from the interaction of cosmic rays with nuclei
in the atmosphere represent the main component of the background.
Atmospheric muons propagate downgoing through the detector
and can be suppressed with an elevation cut, selecting only the upgoing events.
However, some badly reconstructed atmospheric muons
classified as upgoing may remain, and quality cuts are applied
to reduce this contamination to an acceptable level.

In order to establish the criteria used for our neutrino selection,
we have analysed a subsample of data taken by ANTARES after the completion of the
12-line detector, corresponding to a livetime of 70.3 days.
During this period, around 350 upgoing neutrino candidates were reconstructed
and have been compared to a Monte Carlo (MC) simulation of atmospheric muons
and neutrinos using the same livetime.
Downgoing atmospheric muons were simulated with Corsika~\cite{bib:corsika},
and normalized to match the data.
The primary particle flux was
composed of several nuclei according to Ref.~\cite{bib:horandel} and the
QGSJET hadronic model~\cite{bib:qgsjet} was used for the shower development.
Upgoing neutrinos were simulated according to the parameterization
of the atmospheric neutrino flux from Ref.~\cite{bib:bartol}.
Only charged current interactions
of neutrinos and antineutrinos were considered.
The Cherenkov light produced in the vicinity of the ANTARES detector
was propagated taking into account light absorption and scattering in sea
water~\cite{bib:light}. The angular acceptance, quantum efficiency and other
characteristics of the optical modules were taken from Ref.~\cite{bib:om}. 

\begin{figure}[h!]
\centering
\includegraphics[width=0.7\textwidth]{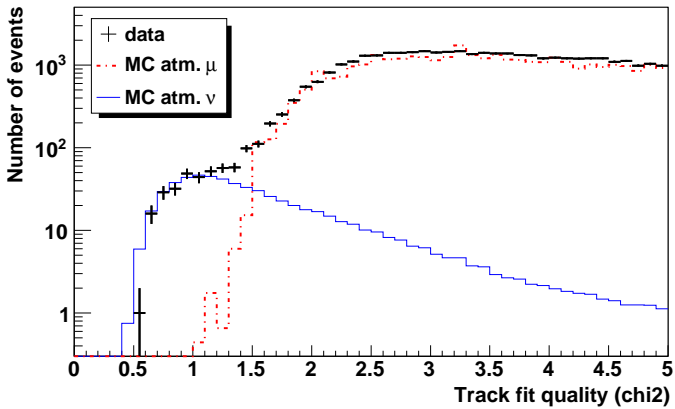}
\caption{Track fit quality ($\chi^2$) distribution for all upgoing events
reconstructed on at least 2 lines. The atmospheric muon Monte Carlo
distribution has been rescaled to match the data.}
\label{fig:FitQuality}
\end{figure}

Figure~\ref{fig:FitQuality} shows the distribution of the track
fit quality ($\chi^2$) of the minimization procedure for
all upgoing events reconstructed with at least two lines. A cut
on the track fit quality is applied to reduce the number of atmospheric
muons reconstructed as upgoing in the final sample. Because the fit
quality is correlated to the number of hits used in the fit,
the selection cut on the fit quality parameter is set to a
different value according to the number of hits used to reconstruct the
event: $\chi^2 \leq 1.3 + [0.04 * (N_{hit}-5)]^2$.
 
\begin{figure}[h!]
\centering
\includegraphics[width=0.7\textwidth]{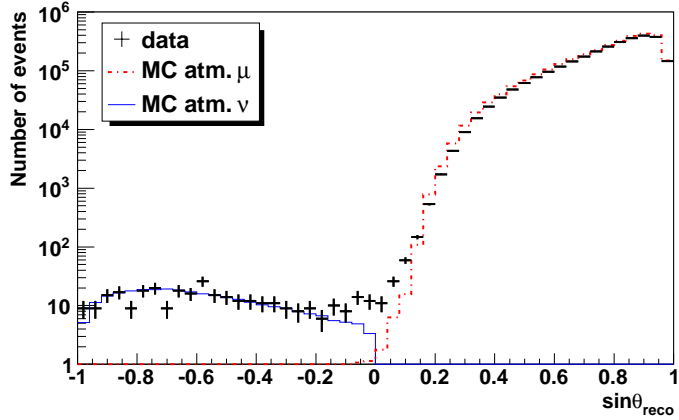}
\caption{Elevation angle distribution after a selection
cut on the fit quality for all events
reconstructed on at least 2 lines.}
\label{fig:CosZenith}
\end{figure}

Figure~\ref{fig:CosZenith} shows the elevation distributions
both for data and simulated atmospheric neutrino and muon samples, after the
cut on the track fit quality.
Atmospheric muons reconstructed as
upgoing are efficiently rejected and a neutrino purity
better than 90\% is achieved.
 
\begin{figure}[h!]
\centering
\includegraphics[width=0.7\textwidth]{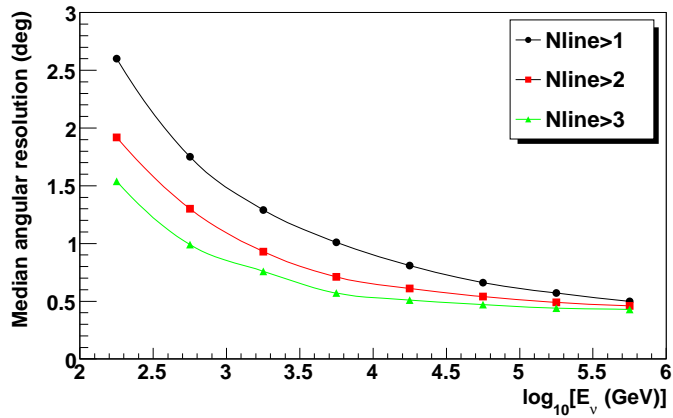}
\caption{Angular resolution as a function of neutrino energy for events
with tracks reconstructed with different number of lines.}
\label{fig:ResoLines}
\end{figure}

Figure~\ref{fig:ResoLines} shows the angular
resolution of the online algorithm
as a function of the neutrino energy for events reconstructed
with different number of lines.
This resolution is defined as the median of
the space angular difference between the incoming neutrino and the
reconstructed neutrino-induced muon.
For neutrinos with an energy higher than a few tens of TeV, an angular
resolution of 0.4 degree is achieved,  
despite of the approximations related to the detector geometry.
For example, the inclination of the ANTARES line for a typical sea current
of 5~cm/s induces a systematic angular deviation of less than 0.2 degree.

\subsubsection{Multi-neutrino trigger}
The typical signature of the transient emission of high energy
neutrinos is a neutrino burst, i.e. a multiplet of neutrino
events originating from the source in a short time window.
A trigger for this event type is implemented as the detection
of two upgoing events reconstructed with at least two lines
in a 15~minutes time window
with a maximum angular difference of 3$^\circ$.
The time window was optimized to include most predictions 
of the neutrino emission by various models for transient sources.
The 3$^\circ$ angular window was selected to match
the convolution of the track reconstruction angular resolution and
the field of view of the robotic optical
telescopes ($\approx 2^\circ \times 2^\circ$).
The accidental coincidence rate due to background events, from two
uncorrelated upgoing atmospheric neutrinos,
is estimated to be $7\times10^{-3}$
coincidences per year with the full ANTARES detector.
With such a small background, the detection of a doublet (triplet) in
ANTARES would have a significance of about 3 (5) sigma.

\subsubsection{High energy event trigger}
Since the neutrino energy spectrum for signal events
is expected to be harder than
for atmospheric neutrinos, a cut on the reconstructed energy
efficiently reduces the atmospheric neutrino background
while most of the signal events are kept.

\begin{figure}[h!]
\centering
\includegraphics[width=0.7\textwidth]{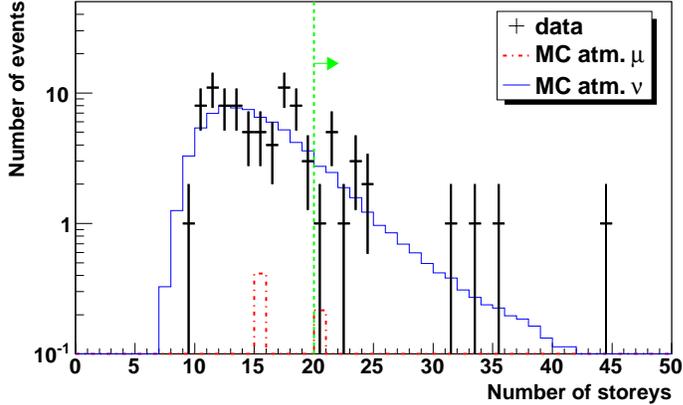}
\caption{Distribution of number of storeys having at least one hit used
in the reconstruction, for data
and Monte Carlo upgoing events reconstructed on at least three lines.
The vertical line indicates the alert selection criterium.}
\label{fig:NumberHits}
\end{figure}

\begin{figure}[h!]
\centering
\includegraphics[width=0.7\textwidth]{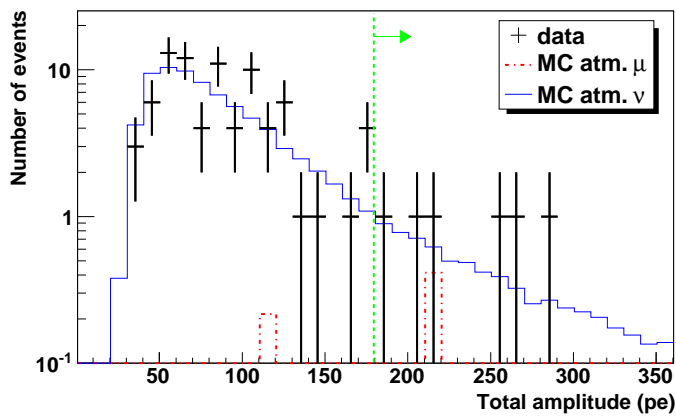}
\caption{Total amplitude distribution in photoelectrons (pe) for
data and Monte Carlo upgoing events reconstructed on at least three lines.
The vertical line indicates the alert selection criterium.}
\label{fig:TotAmplitude}
\end{figure}

The selection of the alert candidates is based on two simple energy estimators:
the number of storeys used in
the track fit and the total amplitude (in photoelectrons) of the hits
in the storeys. 
Figures~\ref{fig:NumberHits} and~\ref{fig:TotAmplitude} show the
distributions of the number of storeys and the amplitude, respectively,
both for data and Monte Carlo samples.

The event selection for the high energy trigger has been tuned on
atmospheric neutrinos in order to obtain a false alarm rate
of about 25 alerts per year. 
This rate was agreed between ANTARES and the optical telescope collaborations.
A requirement of at least 20 storeys on at least three lines and an
amplitude greater than 180 photoelectrons will select 25.7 $\pm$ 0.4
high energy events per year with the full 12 line configuration
of the ANTARES detector.

\subsubsection{Trigger performance}
The performance of these two triggers has been studied using a
neutrino Monte Carlo generated with an $E^{-2}$ energy spectrum.
The TAToO alert criteria select neutrinos of energies above 1 TeV
for the multi-neutrino trigger, and above 10 TeV for the single high
energy trigger.

\begin{figure}[h!]
\centering
\includegraphics[width=0.7\textwidth]{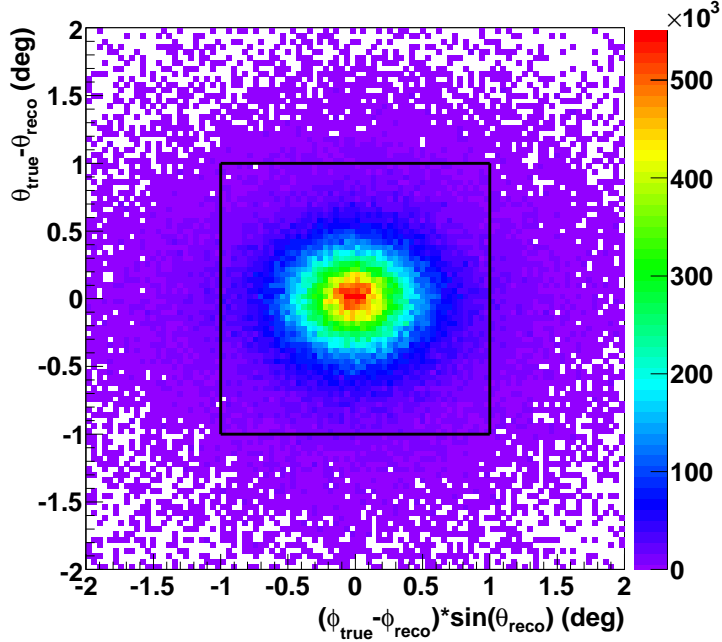}
\caption{Bidimensional point spread function. The black solid-line square
corresponds to a typical robotic telescope
field of view ($\approx 2^\circ \times 2^\circ$).}
\label{fig:ZenithAzimuth}
\end{figure}

\begin{figure}[h!]
\centering
\includegraphics[width=0.7\textwidth]{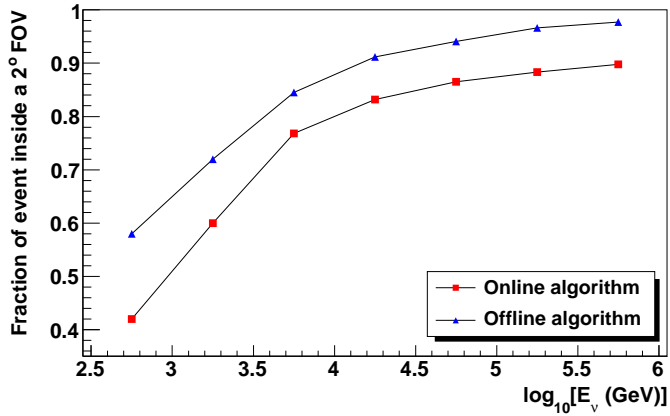}
\caption{Fraction of events inside the field of view
of a typical robotic telescope ($2^\circ \times 2^\circ$) as a function of
the energy of the event, assuming that the tracks originated from the
center of the field of view, for both online and offline reconstruction
algorithms.}
\label{fig:FracEvents}
\end{figure}

The pointing accuracy of the neutrino telescope has two contributions:
the angle between the incoming neutrino and the resulting muon,
and the angle between the muon and the reconstructed trajectory.
The former contribution is due to kinematics: the more energetic the neutrino,
the more co-linear the resulting muon will tend to be with the neutrino
direction. The latter contribution is determined by the performance
of the track reconstruction algorithm.
The bidimensional point spread function
expected for neutrino events selected by the single high energy trigger
is shown in Figure~\ref{fig:ZenithAzimuth}.
As an illustration, $\sim$80\% of the
events are reconstructed within the field of view (FOV) of a typical
robotic telescope.
Figure~\ref{fig:FracEvents} shows the dependence of this fraction on
the neutrino energy.

\section{Determination of the refined direction}
\label{sec:RefinedPosition}
Although the pointing accuracy of the online reconstruction algorithm is
suitable for the field of view of the telescopes used
for the follow-up, the use of the detailed knowledge of
the detector geometry can further improve the determination of the neutrino
direction. For this purpose,
we use the standard ANTARES offline reconstruction algorithm~\cite{bib:AAfit}.

Since the ANTARES lines are not rigid structures,
sea currents can move the top buoy by several
metres, and distort the line positions from a vertical line geometry,
thus affecting the direction of the reconstructed muon trajectory.

In order to achieve the best track reconstruction performance, it is necessary
to monitor the relative positions of all OMs with an accuracy of
better than 20~cm, equivalent to 1~ns timing precision.
In addition, a precise absolute orientation of the whole detector is necessary
to point to individual neutrino sources in the sky.
Two independent monitoring systems are used to attain the required accuracy:
\begin{itemize} 
\item A high frequency long baseline acoustic system giving the 3D position
of hydrophones placed along the line. These positions are obtained by
triangulation from emitters anchored at the bases of the lines.
\item A set of tiltmeter-compass sensors giving the local tilt angles of each
OM storey with respect to the horizontal plane (pitch and roll) as well as its
orientation with respect to the Earth's magnetic field (heading).
\end{itemize}

In order to obtain a quasi-online precise detector geometry (within a
delay of typically few tens of minutes),
the shape of the detector lines is derived
from a model which estimates the mechanical behaviour under the influence of
the sea water flow, obtained from online measurements of the sea
current. The positions of the OMs are then calculated by combining the line
shape with the measurements of the tilt and orientation angles of the storeys
given by the tiltmeter-compass sensors.  

\subsection{The offline likelihood fit}
\label{sec:OfflineReco}
The offline reconstruction algorithm derives the muon track parameters
that maximize a likelihood function built from the difference between the
expected and the measured arrival time of the hits from the Cherenkov
photons emitted along the muon track.
This maximization~\cite{bib:AAfit} takes into account the Cherenkov photons
that scatter in the water and the additional photons that are
generated by secondary particles
(e.g. electromagnetic showers created along the muon trajectory). 

\begin{figure} [htbp]
\centering
\includegraphics[width=0.7\textwidth]{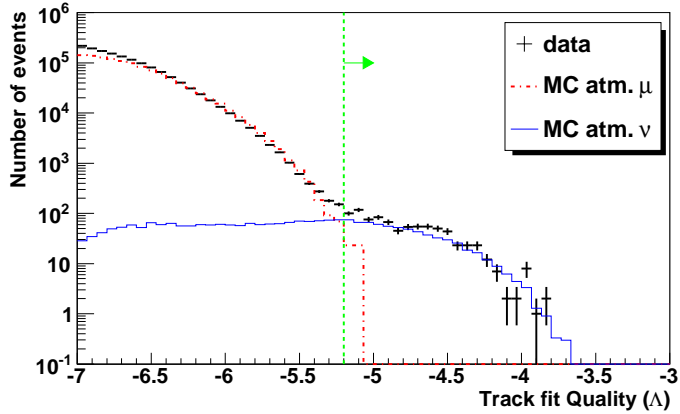}
\caption{Track fit quality ($\Lambda$) distribution for upgoing events
in both data and atmospheric Monte Carlo samples.}
\label{fig:DataMCLambda}
\end{figure}

The value of the log-likelihood per degree of freedom ($\Lambda$)
from the track reconstruction fit is a measure of the track fit quality
and is used to reject badly reconstructed events,
such as atmospheric muons that are mis-reconstructed as upgoing
tracks. The distribution of the variable $\Lambda$ is shown in
Figure~\ref{fig:DataMCLambda} for both data and Monte Carlo events.
Only tracks reconstructed with $\Lambda > -5.2$
are kept for the determination of the refined track direction.

\subsection{Offline reconstruction performance}
\begin{figure} [htbp]
  \centering
\includegraphics[width=0.7\textwidth]{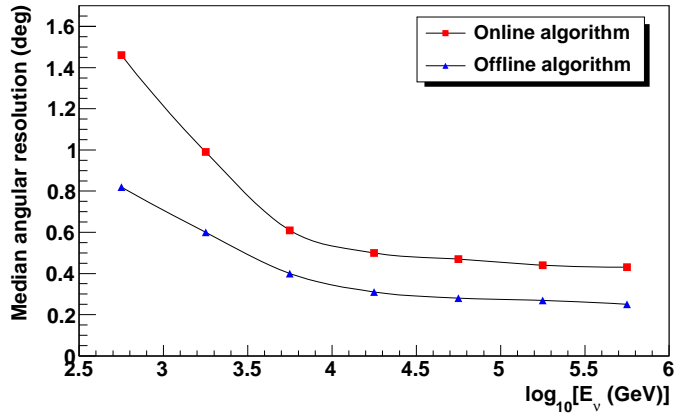}
  \caption{Angular resolution obtained for both online and offline
  reconstructions as a function of the neutrino energy.}
  \label{fig:AngularResolutionAAFitBBFit}
\end{figure}

Figure~\ref{fig:AngularResolutionAAFitBBFit} shows the median
Monte Carlo computed angular
resolution as a function of the neutrino energy for events selected
by the high energy alert trigger, for both online and offline algorithms.
The improvement obtained with the offline algorithm is clearly visible
at low energies.
The angular resolution for neutrino energies above 10 TeV is 0.5$^\circ$
for the online algorithm and 0.35$^\circ$  for the offline algorithm.
As illustrated in Figure 7, $\sim$92\% of the signal events are
reconstructed within the FOV of a typical robotic telescope.

\section{Optical follow-up procedure}
The optical follow-up strategy is based on short-term observations
(for rapidly fading sources) and a long-term follow-up
(mainly for core collapse SNe searches).
Once the alert is sent at time {\it T}$_{0}$, optical images are collected
as soon as possible by the available telescopes.
The fast slewing of the small robotic
telescopes allow the collection of images 5 to 10 seconds after 
the reception of the alert.
The follow-up procedure, which typically extends over 1 month with images
taken each night close to the alert {\it T}$_{0}$,
makes use of the refined direction.
The observations at times later than {\it T}$_{0}$ are
organized as follows: if the refined direction is more than 0.5$^{\circ}$ away
from the initial direction and has a poor offline reconstruction quality, the
follow-up of the alert is cancelled.
Otherwise, the pointing direction of the telescopes is updated with
the refined coordinates and all subsequent images are centered around
that direction.

\section{Summary}
The method used by the ANTARES collaboration
to implement the search for coincidence between high energy neutrinos and
transient sources followed by small robotic telescopes has been presented. 
Of particular importance for this alert system are the ability
to reconstruct online the neutrino direction and to reject efficiently the
background.
With the described ANTARES alert sending capability,
the connected optical telescopes
can start taking images with a latency of the order of one minute,
which will be reduced to about 15~s in the near future. 
The precision of the direction of the alert is much better than one degree.
The quasi-online availability of a refined direction obtained using the
measured geometry of the ANTARES detector further improves the
quality and efficiency of the alert system.

The alert system is operational since early 2009, and as of December 2010,
27 alerts have been sent. After a commissioning phase in 2009, all alerts
had an optical follow-up in 2010.
These numbers are consistent with the expected
trigger rate, after accounting for the duty cycle of the neutrino telescope.

The optical follow-up of neutrino events significantly improves the perspective
for the detection of transient sources. A confirmation by an optical telescope
of a neutrino alert will not only provide information on the nature of
the source but also improve
the precision of the source direction determination in order to
trigger other observatories (for example very large telescopes for
red-shift measurement). The program for the follow-up of ANTARES neutrino
events is already operational with the TAROT and ROTSE telescopes
and results based on analysis of the optical images will be
presented in a forthcoming paper.
This technique could be extended to observations in other wavelength
regimes such as X-ray or radio.
 
\section{Acknowledgments}
%\begin{acknowledgments}
The authors acknowledge the financial support of the funding agencies: 
Centre National de la Recherche Scientifique (CNRS), Commissariat 
\`a l'\'energie atomique et aux \'energies alternatives  (CEA), Agence
National de la Recherche (ANR), Commission Europ\'eenne (FEDER fund 
and Marie Curie Program), R\'egion Alsace (contrat CPER), R\'egion 
Provence-Alpes-C\^ote d'Azur, D\'e\-par\-tement du Var and Ville de 
La Seyne-sur-Mer, France; Bundesministerium f\"ur Bildung und Forschung 
(BMBF), Germany; Istituto Nazionale di Fisica Nucleare (INFN), Italy; 
Stichting voor Fundamenteel Onderzoek der Materie (FOM), Nederlandse 
organisatie voor Wetenschappelijk Onderzoek (NWO), the Netherlands; 
Council of the President of the Russian Federation for young scientists 
and leading scientific schools supporting grants, Russia; National 
Authority for Scientific Research (ANCS), Romania; Ministerio de Ciencia 
e Innovaci\'on (MICINN), Prometeo of Generalitat Valenciana and MultiDark, 
Spain. We also acknowledge the technical support of Ifremer, AIM and 
Foselev Marine for the sea operation and the CC-IN2P3 for the computing facilities.

This work has been financially supported by the GdR PCHE in France. We
want to thank M. Kowalski for discussions on the neutrino triggers and
the organization of the optical follow-up. 
%\end{acknowledgments}

\end{document}